\documentclass[aps,prl,showpacs,preprint,superscriptaddress]{revtex4}
\usepackage{bm}
\usepackage[dvipdfmx]{graphicx}
\usepackage{amsmath}

\begin{document}
\title{Theory of the Inverse Faraday Effect due to the Rashba Spin-Oribt Interactions: Roles of Band Dispersions and Fermi Surfaces}
\author{Yasuhiro Tanaka}
\affiliation{Department of Applied Physics, Waseda University, Okubo, Shinjuku-ku, Tokyo 169-8555, Japan}
\author{Takashi Inoue}
\affiliation{Department of Applied Physics, Waseda University, Okubo, Shinjuku-ku, Tokyo 169-8555, Japan}
\author{Masahito Mochizuki$^{*}$}
\affiliation{Department of Applied Physics, Waseda University, Okubo, Shinjuku-ku, Tokyo 169-8555, Japan}
\begin{abstract}
We theoretically study the inverse Faraday effect, i.e., the optical induction of spin polarization with circularly polarized light, by particularly focusing on effects of band dispersions and Fermi surfaces in crystal systems with the spin-orbit interaction (SOI). By numerically solving the time-dependent Schr\"odinger equation of a tight-binding model with the Rashba-type SOI, we reproduce the light-induced spin polarization proportional to $E_0^2/\omega^3$ where $E_0$ and $\omega$ are the electric-field amplitude and the angular frequency of light, respectively. This optical spin induction is attributed to dynamical magnetoelectric coupling between the light {\it electric} field and the electron spins mediated by the SOI. We elucidate that the magnitude and sign of the induced spin polarization sensitively depend on the electron filling. To understand these results, we construct an analytical theory based on the Floquet theorem. The theory successfully explains the dependencies on $E_0$ and $\omega$ and ascribes the electron-filling dependence to a momentum-dependent effective magnetic field governed by the Fermi-surface geometry. Several candidate materials and experimental conditions relevant to our theory and model parameters are also discussed. Our findings will enable us to engineer the magneto-optical responses of matters via tuning the material parameters.
\end{abstract}
\maketitle

\section{Introduction}
Optical manipulation of electron spins in solids is one of the central issues in condensed matter physics~\cite{Kirilyuk10}, which has been intensively studied both experimentally~\cite{Beaurepaire96,Koshihara_PRL97,Koopmans00,Tudosa04,JuG04,Kimel_NAT05,Hansteen05,Stanciu07a,Stanciu07b,Bigot09,Kampfrath_NAP11,Nishizawa_PNAS17,Miyamoto_SR18} and theoretically~\cite{Takayoshi_PRB14,Takayoshi_PRB14-2,Sato_PRL16,Mochizuki_APL18,Kozin_PRB18,Pitaevskii61,Pershan66,Hertel05,Battiato_PRB14,Zhang_PLS19}. The usage of light as a means for the spin manipulation has numerous advantages over other methods. These advantages include (1) ultrafast response speeds on picosecond or shorter time-scales, (2) contactless operations free from frictional wear, and (3) enhanced highly efficient responses if resonant excitations are exploited. The spin induction by circularly polarized light, the so-called inverse Faraday effect, was first proposed by Pitaevskii in 1961 based on a phenomenological theory for a continuum medium, which predicted the light-induced spin polarization proportional to $E_0^2$ with $E_0$ being the electric-field amplitude of light~\cite{Pitaevskii61}. Subsequently, the microscopic theory for an isolated ion was proposed by Pershan and coworkers in 1966 based on perturbation expansions with respect to the light electric field, which predicted the induced spin polarization inversely proportional to $\omega^3$ with $\omega$ being the angular frequency of light~\cite{Pershan66}. 

A recent theoretical study proved that the spin-orbit interaction (SOI) offers highly efficient ways of optical spin induction with circularly polarized light~\cite{Mochizuki_APL18}. It was theoretically demonstrated that in electron systems with SOI, the rotating {\it electric} field of the incident light (instead of the rotating {\it magnetic} field) can induce spin polarization much more efficiently because the coupling energy between the light electric field and the electron charges is several orders of magnitude larger than that between the light magnetic field and the electron spins. Here the rotating motion of the electrons induced by the rotating electric field of light is converted to a strong rotating magnetic field, which gives rise to an effective static magnetic field normal to the plane of light polarization. This mechanism is in contrast to that proposed for localized spins in Mott insulators, where the rotating magnetic field of light induces spin polarization via Zeeman coupling~\cite{Takayoshi_PRB14,Takayoshi_PRB14-2}. 
However, the work is based on a phenomenological approach where the time evolution of a Gaussian wave packet was simulated by solving the time-dependent Schr\"odinger equation and, thereby, missed the effects of band-structure formation in crystalline materials.

\begin{figure}[t]
\includegraphics[width=10cm
]{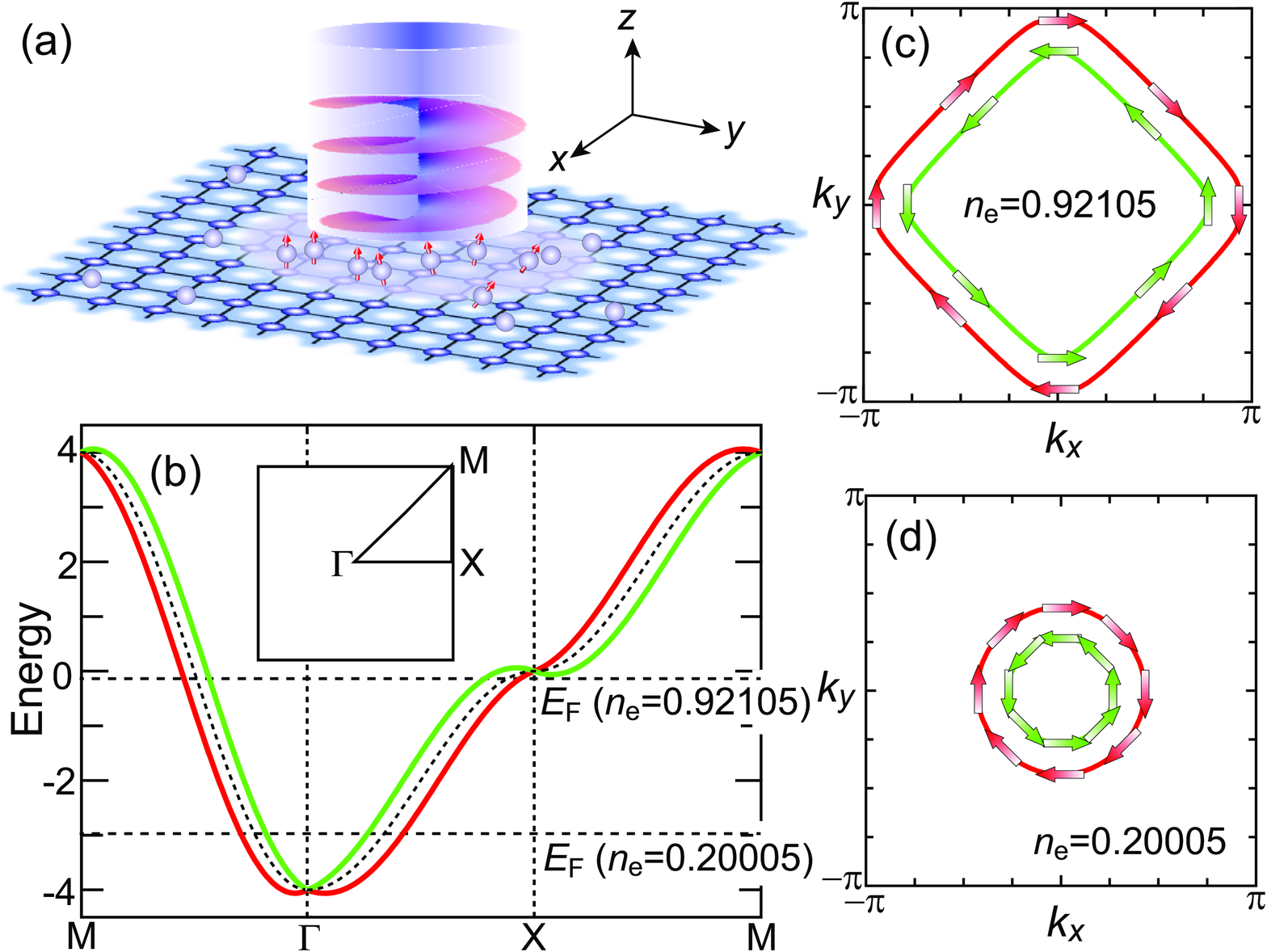}
\caption{(a) Schematic illustration of the optical induction of spin polarization with circularly polarized light for an electron system with the SOI. (b) Band dispersions (${\bm A}={\bm 0}$ and $B_0=0$) in the presence of Rashba-type SOI with $\alpha_{\rm R}=0.5$ (solid lines) and those in the absence of SOI (dashed line). Fermi levels for electron fillings of $n_{\rm e}$=0.92105 and $n_{\rm e}$=0.20005 are shown by horizontal dashed lines. (c), (d) Fermi surfaces for different electron fillings of (c) $n_{\rm e}$=0.92105 and (d) $n_{\rm e}$=0.20005. Spin orientations at several points on the Fermi surfaces are indicated by arrows.}
\label{Fig1}
\end{figure}

In this paper, we theoretically investigate the inverse Faraday effect in an electron system with the Rashba-type SOI [Fig.~\ref{Fig1}(a)] by particularly focusing on the effects of band formation in crystalline materials. We first perform numerical simulations by solving the time-dependent Schr\"odinger equation of a tight-binding model with the Rashba SOI. The simulated spatiotemporal spin dynamics show that the light indeed induces spin polarization perpendicular to its polarization plane, magnitude of which is proportional to $E_0^2/\omega^3$. We also discover sensitive electron-filling dependence of magnitude and sign of the induced spin polarization. In order to understand these simulation results, we construct a theory based on the Floquet theorem~\cite{Kitagawa_PRB11,Lindner_NATP11,Eckardt_NJP15}, which describes nonequilibrium steady states under a continuous time-periodic excitation. The theory show that the rotating electric field of light is converted to an effective static magnetic field perpendicular to the plane of light polarization by the SOI, which eventually reproduces the spin polarization proportional to $E_0^2/\omega^3$. This SOI-mediated effective magnetic field turns out to have remarkable momentum dependence. The observed sensitive filling dependence of the spin polarization can be explained by this momentum dependence governed by the Fermi-surface geometry. Our microscopic theory for crystalline materials uncovered important roles of the band dispersions and the Fermi surfaces for the inverse Faraday effect, which have been missed so far. These findings provide us with a firm basis to design the magneto-optical responses of solids via tuning the material parameters.

\section{Numerical simulations}
\begin{table}
\begin{tabular}{ccc}\hline
Quantity & Dimensionless quantity & Corresponding values \\ \hline
Frequency & $\omega=\hbar\tilde{\omega}/t$=1 & $\tilde{\omega}$=242 THz\\ 
Light $E$ field & $E_0=ea\tilde{E}_0/t$=1 & $\tilde{E}_0$=20 MV/cm\\ 
Light $B$ field & $B_0=g\mu_{\rm B}\tilde{E}_0/(2ct)$ & $\tilde{B}_0$=6.67 T\\ 
  & \hspace{0.4cm}$=3.86 \times 10^{-4}$ & \\ 
Time & $\tau=\tilde{\tau}t/\hbar$=1 & $\tilde{\tau}$=0.66 fs\\ \hline
\end{tabular}
\caption{Unit conversion table for $t$=1 eV and $a$=5 \AA. The symbols $\omega$, $E_0$, $B_0$ and $\tau$ denote dimensionless variables used in this paper for light frequency, light electric field, light magnetic field, and time, respectively, when $t$ and $a$ are taken as the units of energy and length with the natural units $e$=$\hbar$=$c$=1. The symbols $\tilde{\omega}$, $\tilde{E}_0$, $\tilde{B}_0$, and $\tilde{\tau}$ are variables for the real quantities. Note that the relation $\tilde{B}_0=\tilde{E}_0/c$ holds for electromagnetic waves, which gives $B_0=3.86\times 10^{-4}$ ($\tilde{B}_0$=6.67 T) when $E_0=1$ ($\tilde{E}_0$=20 MV/cm).}
\label{uc-table}
\end{table}

We consider a time-dependent Hamiltonian, which is composed of three terms:
\begin{equation}
{\mathcal H}(\tau)
={\mathcal H}_0(\tau)+{\mathcal H}_{\rm so}(\tau)+{\mathcal H}_{\rm Zeeman}(\tau).
\label{eq-Htot}
\end{equation}
The first term ${\mathcal H}_0(\tau)$ denotes a tight-binding model for electrons on a square lattice irradiated with a time-dependent electromagnetic field. This term is given by,
\begin{eqnarray}
{\mathcal H}_0(\tau)=\sum_{<i,j>}t \exp\left[ -i\bm A(\tau) \cdot \bm e_{ij}
\right] c^\dagger_i c_j.
\label{eq-H0A}
\end{eqnarray}
where the symbol $t$ represents the nearest neighbor transfer integrals, the vector $\bm e_{ij}$ denotes the unit directional vector connecting the adjacent $i$th and $j$th sites, and the vector $\bm A(\tau)$ rereresents the vector potential generated by the light electromagnetic field as will be explained shortly. Note that we adopt natural units $e$=$\hbar$=$c$=1 and take the transfer integral $t$ and the lattice constant $a$ as the units of energy and length, respectively. After the Fourier transformation, this term is rewritten in the momentum representation as,
\begin{equation}
{\mathcal H}_0(\tau)=\sum_{\bm k,\sigma}\varepsilon_{\bm k,A}
c^\dagger_{\bm k\sigma}c_{\bm k\sigma}.
\label{eq-H0}
\end{equation}
The second term ${\mathcal H}_{\rm so}(\tau)$ describes the SOI~\cite{Mireles_PRB01,Alex_PRB10}. In the momentum space, the second term is given by,
\begin{eqnarray}
{\mathcal H}_{\rm so}(\tau)
=-\alpha_{\rm R}\sum_{\bm k}\left[
\sin(k_x+A_x)c^\dagger_{\bm k\sigma}(\sigma_y)_{\sigma\sigma'}c_{\bm k\sigma'}
-\sin(k_y+A_y)c^\dagger_{\bm k\sigma}(\sigma_x)_{\sigma\sigma'}c_{\bm k\sigma'}
\right].
\label{eq-Hso}
\end{eqnarray}
We set the lattice constant $a$ as the units of length ($a$=1) hereafter. The third term ${\mathcal H}_{\rm Zeeman}(\tau)$ describes the Zeeman coupling between the electron spins and the rotating light magnetic field $\bm B(\tau)=(B_x(\tau), B_y(\tau), 0)$, which is given by,
\begin{eqnarray}
{\mathcal H}_{\rm Zeeman}(\tau)=
\sum_{\bm k}\left[
 B_x(\tau) c^\dagger_{\bm k\sigma}(\sigma_x)_{\sigma\sigma'}c_{\bm k\sigma'}
+B_y(\tau) c^\dagger_{\bm k\sigma}(\sigma_y)_{\sigma\sigma'}c_{\bm k\sigma'}
\right].
\label{eq-Hzeeman}
\end{eqnarray}
Here $c^\dagger_{k\sigma}$ ($c_{k\sigma}$) denotes the creation (annihilation) operator for an electron with wave vector $\bm k$ and spin $\sigma$ ($=\uparrow, \downarrow$). The symbols $\sigma_{\alpha}$ ($\alpha=x,y,z$) and $\alpha_{\rm R}$ represent the Pauli matrices and strength of the Rashba SOI, respectively. Note that the laser irradiation can induce a weak time-periodic variation of the SOI through temporally modulating the atomic and electronic structures. However, the strength of SOI is predominantly determined by the crystal structure with broken spatial inversion symmetry, and the influence of the laser-induced modulations of the electronic and atomic states are expected to be negligibly weak particularly in the present perturbational regime. Thus we adop a steady Rashba parameter $\alpha_{\rm R}$ throughout the present study.

The SOI manifests itself in systems without spatial inversion symmetry~\cite{Winkler_Book,Rashba_SP60,Dress_PR55,Manchon_NAM15,Bercioux_RPP15}. The Rashba SOI, for example, becomes active in semiconductor heterostructures, magnetic multilayer systems, and surfaces of magnetic thin films~\cite{Datta_APL90,Kohda_JPSJ08,Nitta_PRL97,Koo_SCI09,Sugahara_APL04,Sasaki_PRA14,Edelstein_SSC90,Kato_PRL04,Jungwirth_NAM12,Kunihashi_NAC16}, whereas the Dresselhaus SOI appears in bulk III-V semiconductors, e.g., GaAs and InAs, because of the absence of inversion symmetry in their crystal structures~\cite{Miller_PRL03,Ganichev_PRL04}. In the present study, we consider the Rashba-type SOI for ${\mathcal H}_{\rm so}(\tau)$, but the results do not alter even qualitatively if we use the Dresselhaus-type SOI. The band dispersion $\varepsilon_{\bm k,A}$ is given in the form,
\begin{equation}
\varepsilon_{\bm k,A}=-2t[\cos(k_x+A_x)+\cos(k_y+A_y)],
\end{equation}
which is obtained by the Fourier transformation of the tight-binding Hamiltonian $\mathcal{H}_0(\tau)$ in Eq.~(\ref{eq-H0A}).

The coupling between the electrons and the light electric field is incorporated via the Peierls substitution. We choose the temporal gauge in which the scalar potential $\phi$ is set to be zero. In this case, the time-dependent vector potential, $\bm A(\tau)$, is given in the form
\begin{equation}
{\bm A}(\tau)=-\int^{\tau}_{0}{\bm E}(\tau')d\tau'.
\label{eq-A}
\end{equation}
The time-dependent light electric field is given by
\begin{equation}
\bm E(\tau)=E_0\beta(\tau)(\cos\omega \tau, \sin\omega \tau),
\label{eq-E}
\end{equation}
with amplitude $E_0$ and angular frequency $\omega$. Then the magnetic field $\bm B(\tau)$ for left-handed circularly polarized light is given in the form,
\begin{equation}
\bm B(\tau)=B_0\beta(\tau)(\sin\omega \tau, -\cos\omega \tau).
\label{eq-B}
\end{equation}
Using this expression, the Zeeman-coupling term ${\mathcal H}_{\rm Zeeman}(\tau)$ is rewritten in the form,
\begin{eqnarray}
{\mathcal H}_{\rm Zeeman}(\tau)=B_0\beta(\tau)
\sum_{\bm k}\left[
 (\sin\omega\tau) c^\dagger_{\bm k\sigma}(\sigma_x)_{\sigma\sigma'}c_{\bm k\sigma'}
-(\cos\omega\tau) c^\dagger_{\bm k\sigma}(\sigma_y)_{\sigma\sigma'}c_{\bm k\sigma'}
\right].
\label{eq-HZeeman}
\end{eqnarray}
Table \ref{uc-table} gives unit conversions for $\omega$, $E_0$, $B_0$, and time $\tau$ when $t$=1 eV and $a$=5 \AA, which are typical values for semiconductors. We introduce a factor $\beta(\tau)=1-e^{-\tau^2/\tau^2_d}$ that causes the external field to rise gradually so as to avoid impact forces on the spins and the resulting artificial spin oscillations in the photoinduced dynamics~\cite{Mochizuki_APL18}.

The matrix representation of ${\mathcal H}(\tau)$ is given by
\begin{equation}
{\mathcal H}(\tau)=\sum_k(c^\dagger_{\bm k\uparrow},c^\dagger_{\bm k\downarrow})
\begin{pmatrix}
\varepsilon_{\bm k,A} & \gamma_{\bm k,A}+b_\omega \\
\gamma^{\ast}_{\bm k,A}+b^{\ast}_\omega & \varepsilon_{\bm k,A}
\end{pmatrix}
\begin{pmatrix}
c_{\bm k\uparrow}  \\
c_{\bm k\downarrow}  
\end{pmatrix},
\label{eq-Hm}
\end{equation}
where
\begin{equation}
\gamma_{\bm k,A}=\alpha_{\rm R}[i\sin(k_x+A_x)+\sin(k_y+A_y)],
\end{equation}
and
\begin{equation}
b_\omega=B_0\beta(\tau) \left[\sin\omega\tau+i\cos\omega\tau\right].
\end{equation}
The time evolution of the system is simulated using the time-dependent Schr\"odinger equation;
\begin{equation}
i\partial_\tau|\Psi_{\bm k,\nu}(\tau)\rangle ={\mathcal H}(\tau)|\Psi_{\bm k,\nu}(\tau)\rangle, 
\label{eq-Scheq}
\end{equation}
where $|\Psi_{\bm k,\nu}(\tau)\rangle$ is the $\nu$th ($\nu=1,2$) one-particle state with wave vector $k$. We numerically solve a discretized equation,
\begin{equation}
|\Psi_{\bm k,\nu}(\tau+\Delta \tau)\rangle = \exp[-i\Delta \tau {\mathcal H}(\tau+\Delta \tau/2)]
|\Psi_{\bm k,\nu}(\tau)\rangle, 
\label{eq-Scheq2}
\end{equation}
where the accuracy of the obtained $|\Psi_{\bm k,\nu}(\tau)\rangle$ is within an error of the order of $(\Delta \tau)^3$~\cite{Terai_PTPS93,Kuwabara_JPSJ95,Tanaka_JPSJ10}. In the present study, we adopt $\Delta \tau=0.01$ which guarantees sufficient accuracy of the numerical simulations. We use a system of $N=L\times L$ with $L=200$ and impose periodic boundary conditions. 

The energy dispersion relations $E(\bm k)=\varepsilon_{\bm k}\pm |\gamma_{\bm k}|$ in the absence of light irradiation ($\bm A=\bm 0$ and $B_0$=0) are shown in Fig.~\ref{Fig1}(b) for $\alpha_{\rm R}=0.5$. Here $\varepsilon_{\bm k}$ and $\gamma_{\bm k}$ denote $\varepsilon_{\bm k,A}$ and $\gamma_{\bm k,A}$ with ${\bm A}={\bm 0}$, respectively. The spin degeneracy of the bands is lifted by the SOI. The Fermi surfaces for $n_{\rm e}$=0.92105 and $n_{\rm e}$=0.20005 are shown in Figs.~\ref{Fig1}(c) and \ref{Fig1}(d), respectively. The spin orientations at several points on the Fermi surfaces are indicated by arrows. We find that the Fermi surfaces are rounded squares for the higher electron filling of $n_{\rm e}$=0.92105, whereas they are circular for the lower electron filling of $n_{\rm e}$=0.20005.

\begin{figure}[t]
\includegraphics[
width=8cm]{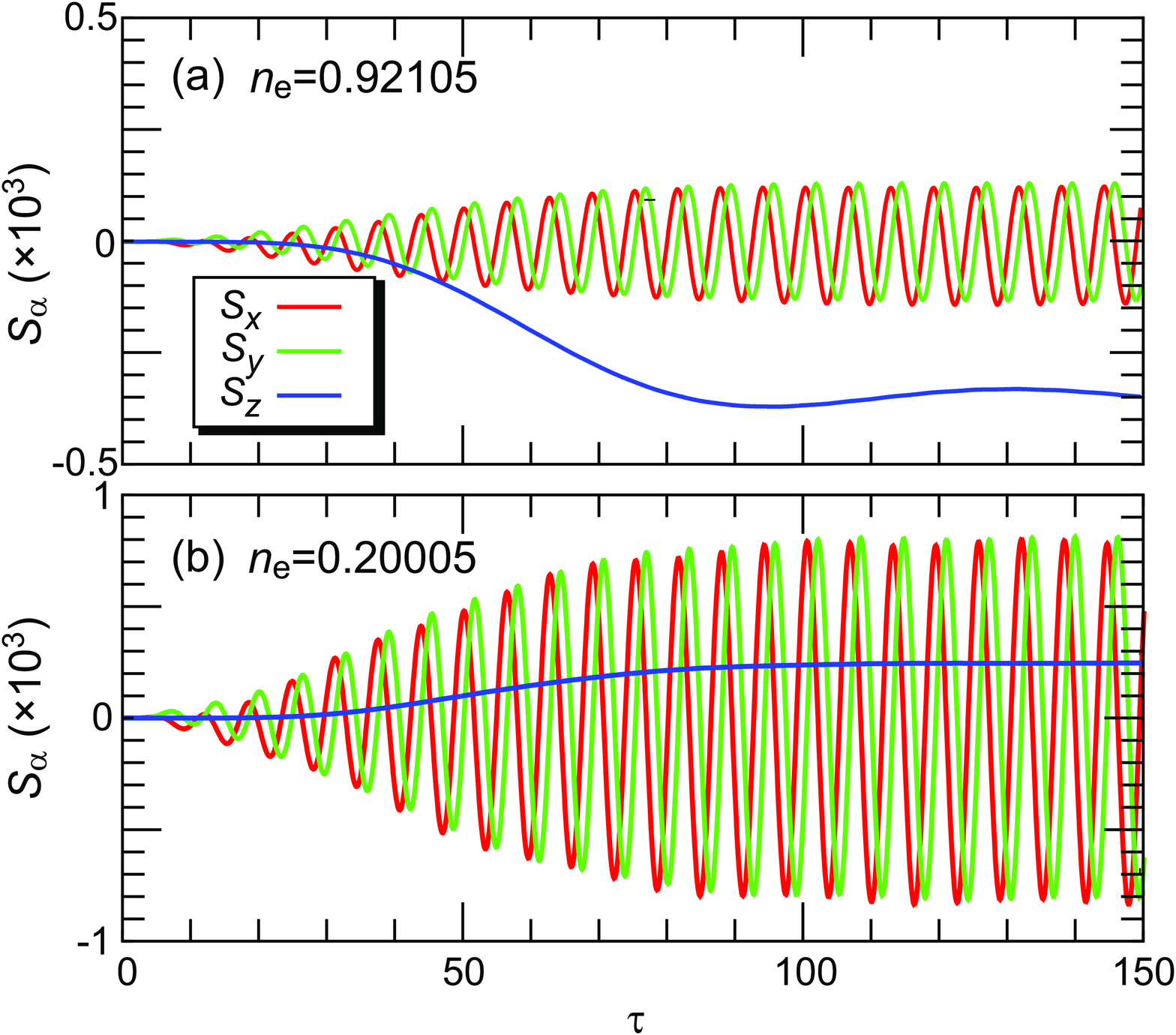}
\caption{Simulated time evolutions of the net spin $\bm S(\tau)=(S_x(\tau), S_y(\tau), S_z(\tau))$ for electron fillings of (a) $n_{\rm e}$=0.92105 and (b) $n_{\rm e}$=0.20005 when $\alpha_{\rm R}=0.1$, $\omega=1$, $E_0=0.1$, and $\tau_d=50$. See Table \ref{uc-table} for unit conversions.}
\label{Fig2}
\end{figure}
In Figs.~\ref{Fig2}(a) and \ref{Fig2}(b), we show simulated time evolutions of the net spin, $\bm S(\tau)$=$(S_x(\tau), S_y(\tau), S_z(\tau))$, for electron fillings of $n_{\rm e}$=0.92105 and $n_{\rm e}$=0.20005, respectively. They are calculated by
\begin{eqnarray}
S_{\alpha}=\frac{1}{2N_k}\sum_{\bm k}\langle c^\dagger_{\bm k\sigma}(\sigma_{\alpha})_{\sigma \sigma'} c_{\bm k\sigma'}\rangle
\end{eqnarray}
where $N_k$ is the number of $k$ points in the area between the outer and inner Fermi surfaces in the presence of the SOI. We used $\alpha_{\rm R}=0.1$, $\omega=1$, $E_0=0.1$ and $\tau_d=50$ for the simulations. In both cases, the in-plane components $S_x(\tau)$ and $S_y(\tau)$ show sinusoidal oscillations around zero whose frequency coincides with the light frequency $\omega$. The phase of $S_x(\tau)$ advances by $\pi/2$ compared with that of $S_y(\tau)$, indicating that the in-plane component of total spin rotates in an anticlockwise direction. On the contrary, the out-of-plane component $S_z(\tau)$ exhibits saturation. For $n_{\rm e}$=0.92105, it decreases gradually in a transient process for $\tau<100$ and nearly saturates to a finite negative value after sufficient duration ($\tau>100$). For $n_{\rm e}$=0.20005, it converges to a finite positive value for $\tau>80$. These results show that the spin polarization appears perpendicular to the plane, and its magnitude and sign depend sensitively on the electron filling.

\begin{figure}
\includegraphics[
width=8cm]{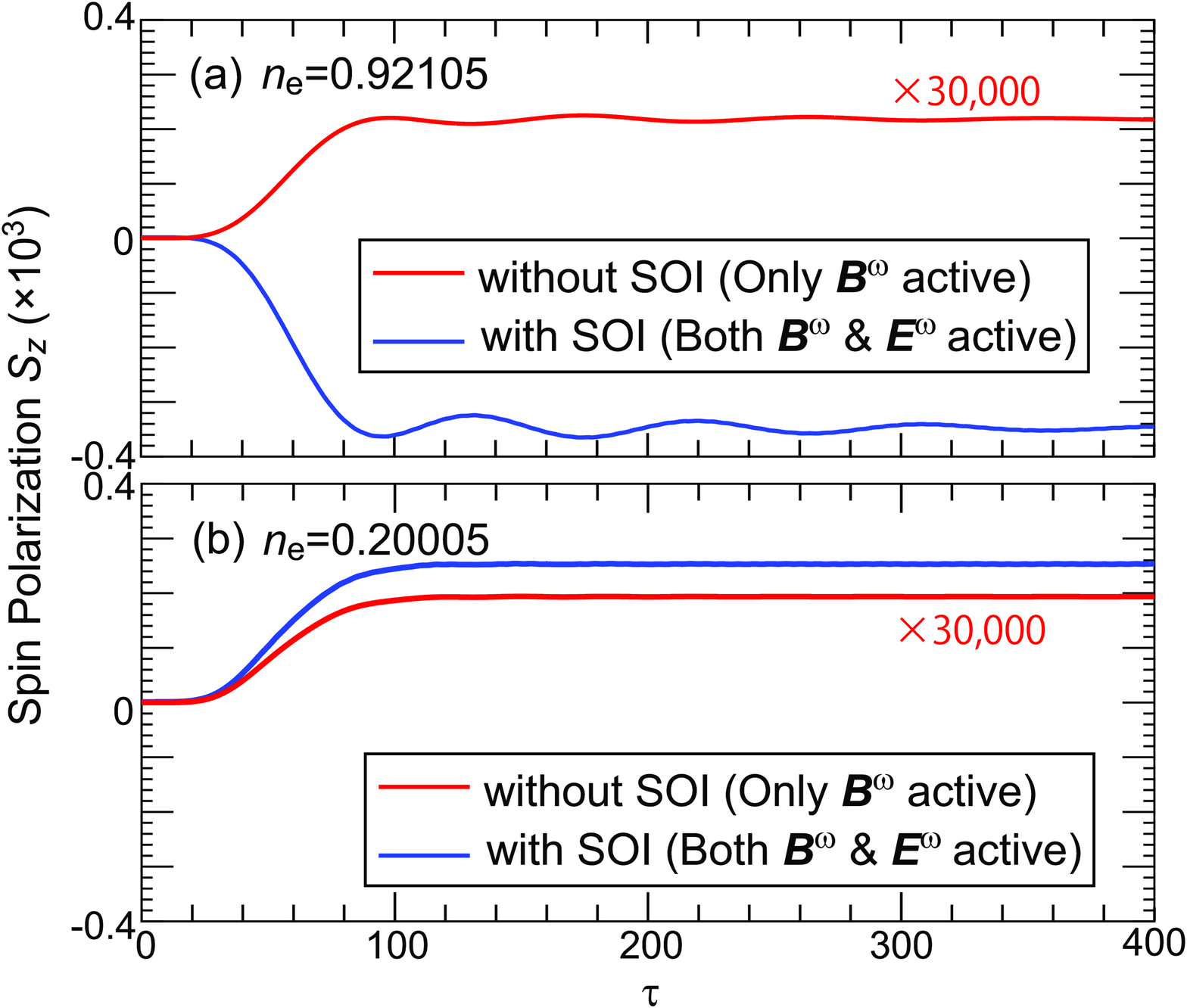}
\caption{(a) Simulated time evolutions of the photoinduced spin polarization $S_z(\tau)$ in the presence ($\alpha_{\rm R}=0.1$) and absence ($\alpha_{\rm R}=0$) of the Rashba SOI for the electron filling of $n_{\rm e}$=0.92105. (b) Those for the electron filling of $n_{\rm e}$=0.20005. In the presence of the Rashba SOI, the electron spins are activated both by the magnetic-field ($\bm B^\omega$) and the electric-field ($\bm E^\omega$) components of light, whereas they are activated only by the $\bm B^\omega$ component of light in the absence of the Rashba SOI. The simulations are performed for $\omega=1$, $E_0=0.1$, and $\tau_d=50$. See Table \ref{uc-table} for unit conversions.}
\label{Fig3}
\end{figure}
In fact, this optical spin-polarization induction becomes highly efficient in the presence of the SOI. To demonstrate crucial roles of the SOI, we perform the simulations for both the cases with and without the Rashba SOI. In Figs.~\ref{Fig3}(a) and (b), we plot the simulated time profiles of the photoinduced spin polarization $S_z(\tau)$ in the presence ($\alpha_{\rm R}=0.1$) and absence ($\alpha_{\rm R}=0$) of the Rashba SOI for different electron fillings of (a) $n_{\rm e}$=0.92105 and (b) $n_{\rm e}$=0.20005. For both electron fillings, the case with the Rashba SOI exhibits much larger spin polarization than the case without the Rashba SOI. The values of $S_z(\tau)$ for the system without the Rashba SOI are negligibly small so that we need to multiply by 30,000 to make them visible in the present plot scale. The substantial difference between the two cases is that the electric-field component of light couple to the electron spins when the SOI is present, whereas it cannot when the SOI is absent.

\begin{figure*}[t]
\includegraphics[
width=15cm]{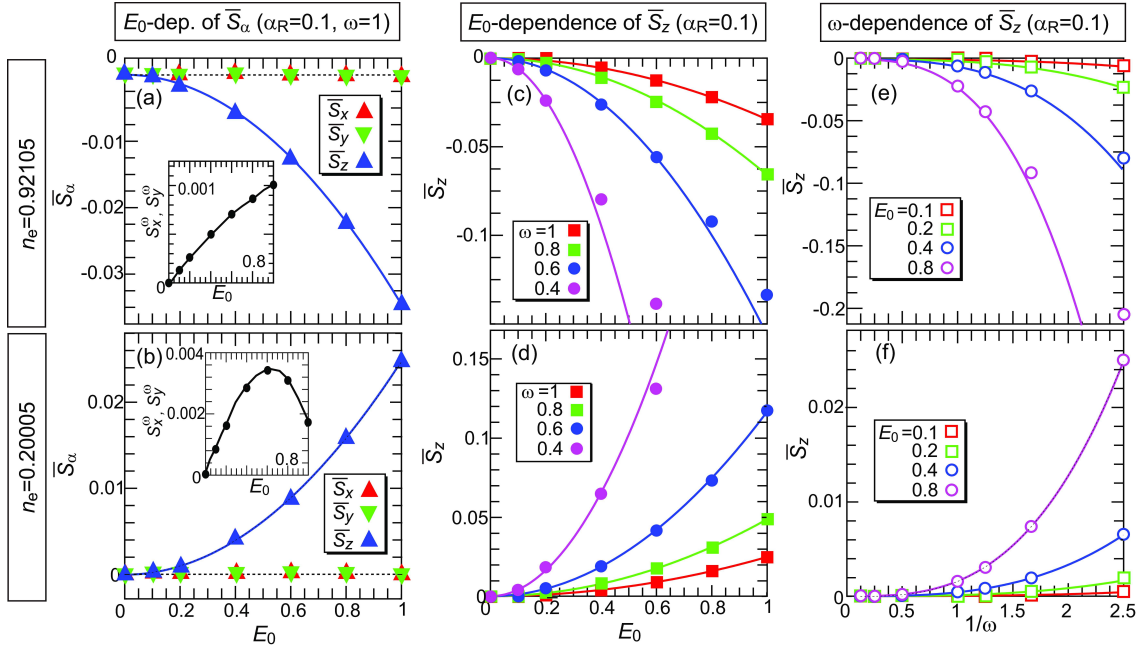}
\caption{(a), (b) Time averages of in-plane components of the net spin $\overline{S_x}$ and $\overline{S_y}$ and saturated spin polarization $\overline{S_z}$ after sufficient duration as functions of the light electric field $E_0$ when $\omega=1$ for different electron fillings of (a) $n_{\rm e}$=0.92105 and (b) $n_{\rm e}$=0.20005. Here the dashed lines indicate zero, and the insets show oscillation amplitudes of $S_x(\tau)$ and $S_y(\tau)$. (c), (d) Saturated spin polarizations $\overline{S_z}$ as functions of $E_0$ for various values of $\omega$ for (c) $n_{\rm e}$=0.92105 and (d) $n_{\rm e}$=0.20005. (e), (f) Saturated spin polarizations $\overline{S_z}$ as functions of $1/\omega$ for various values of $E_0$ for (e) $n_{\rm e}$=0.92105 and (f) $n_{\rm e}$=0.20005. Solid lines in (a)-(d) are fitting curves with $\overline{S_z} \propto E_0^2$, whereas those in (e) and (f) are fitting curves with $\overline{S_z} \propto 1/\omega^3$. For unit conversions, see Table \ref{uc-table}.}
\label{Fig4}
\end{figure*}
The simulation data in Fig.~\ref{Fig3} indicate that the electric-field component of light (instead of the magnetic-field component) dominates the optical spin induction. This is because an energy scale of the coupling between the light electric field and the electron charges is much larger than that of the Zeeman coupling between the light magnetic field and the electron spins. Note that the relation $E^\omega/B^\omega=c$ holds between the electric-field amplitude $E^\omega$ and the magnetic-field amplitude $B^\omega$ for electromagnetic waves with $c$ being the speed of light. This relation indicates that $B^\omega$ is only $\sim 0.3$ T even for a relatively intense laser with $E^\omega$=1 MV/cm. An energy scale of the Zeeman interaction between the magnetic field $B^\omega$=0.3 T and an electron spin ($S=\hbar/2$) is evaluated to be only $\sim 2 \times 10^{-5}$ eV, whereas that of the Coulomb interaction between the electric field $E^\omega$=1 MV/cm and an electron charge $e$ is evaluated to be $\sim 5 \times 10^{-2}$ eV if we assume a typical lattice constant of $a$=5 \AA. Namely the latter energy scale is more than three orders of magnitude larger, and we can exploit this strong coupling between the $E^\omega$ field and the electrons in the Rashba-SOI system.

In Figs.~\ref{Fig4}(a) and \ref{Fig4}(b), we show calculated $E_0$ dependence of the time averages of net spin, $\overline{S_\alpha}$ ($\alpha$=$x$, $y$, $z$), for electron fillings of $n_{\rm e}$=0.92105 and $n_{\rm e}$=0.20005, respectively. Here the light frequency is fixed at $\omega$=1. The time averages are taken over a time period of $300<\tau<400$, in which the out-of-plane spin component $S_z(\tau)$ is nearly saturated to be a constant value and thus its time average $\overline{S_z}$ almost coincides with the saturation value. We find that a finite spin polarization $\overline{S_z}$ proportional to $E_0^2$ appears, whereas the in-plane components $\overline{S_x}$ and $\overline{S_y}$ are always zero. Importantly, the sign of the induced $\overline{S_z}$ differs depending on the electron filling: it is negative for the higher electron filling of $n_{\rm e}$=0.92105, whereas it is positive for the lower electron filling of $n_{\rm e}$=0.20005. It is also noteworthy that, as seen in the insets of Figs.~\ref{Fig4}(a) and \ref{Fig4}(b), the oscillation amplitudes of $S_x$ and $S_y$ increase almost linearly with increasing $E_0$ for $n_{\rm e}$=0.92105, whereas it is not monotonic for $n_{\rm e}$=0.20005.

In Figs.~\ref{Fig4}(c) and \ref{Fig4}(d), the calculated spin polarization $\overline{S_z}$ for several values of $\omega$ are plotted as functions of the amplitude of light electric field $E_0$ for different electron fillings, which show that the quadratic $E_0$ dependence of $\overline{S_z}$ and the filling dependence of its sign hold even when $\omega$ is varied. It should, however, be noted that deviation from the quadratic $E_0$ dependence appears in the region of large $E_0$ when $\omega$ is small. We also find that $|\overline{S_z}|$ takes larger values for s smaller $\omega$. increases with decreasing $\omega$. 

The increase of $\overline{S_z}$ with decreasing $\omega$ can be clearly seen in the calculated $1/\omega$ dependence of $\overline{S_z}$ plotted in Figs.~\ref{Fig4}(e) and \ref{Fig4}(f). We find that the induced spin polarization $\overline{S_z}$ increases with increasing $1/\omega$. Moreover, this $\omega$ dependence turns out to be well fitted by $\overline{S_z} \propto 1/\omega^3$, although deviations from the fitting again appear when both $1/\omega$ and $E_0$ are large. 

\begin{figure}
\includegraphics[
width=9cm]{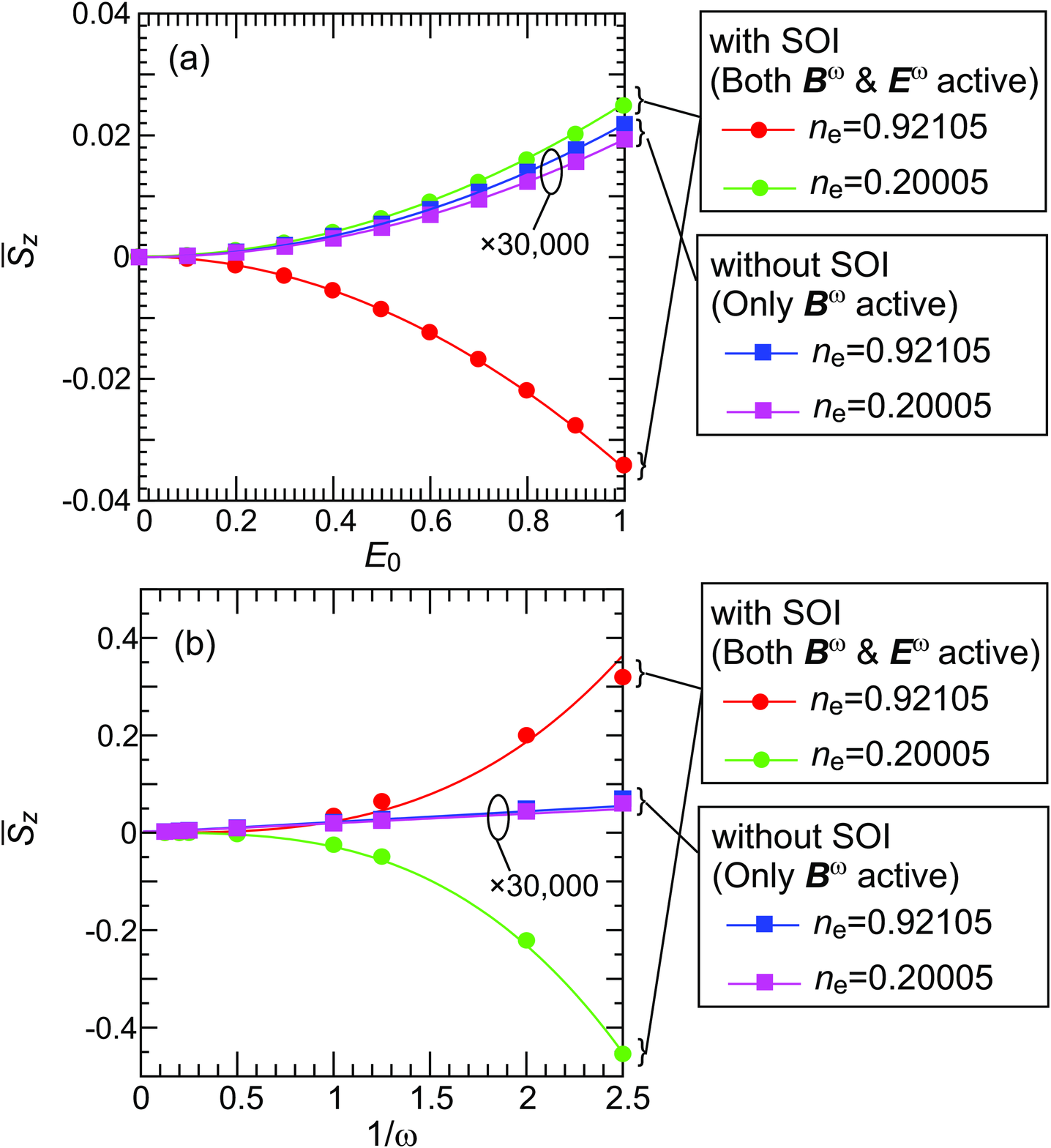}
\caption{(a) $E_0$ dependence and (b) $1/\omega$ dependence of the induced spin polarization $\overline{S_z}$ for electron fillings of $n_e$=0.92105 and $n_e$=0.20005 in the cases with and without the Rashba SOI where $E_0$ and $\omega$ are the amplitude of electric field and the frequency of light. The parameters are set to be $\alpha_{\rm R}$=0.1, $\omega$=1 for (a) and $\alpha_{\rm R}$=0.1, $E_0$=1 for (b). Comparisons between the cases with and without Rashba SOI show that the Rashba SOI significantly enhances the inverse Faraday effect. For unit conversions, see Table \ref{uc-table}.}
\label{Fig5}
\end{figure}
To demonstrate the critical roles of SOI, we compare the spin polarizations $\overline{S_z}$ in the cases with and without Rashba SOI. In Figs.~\ref{Fig5}(a) and (b), we show the calculated (a) $E_0$ dependence and (b) $1/\omega$ dependence of $\overline{S_z}$ for $n_e$=0.92105 and $n_e$=0.20005. The parameters used for the calculations are $\alpha_{\rm R}$=0.1 and $\omega$=1 for (a), whereas $\alpha_{\rm R}$=0.1 and $E_0$=1 for (b). Note that in the presence of SOI, both the electric-field ($\bm E^\omega$) and the magnetic-field ($\bm B^\omega$) components of light can couple to the electron spins and contribute to the spin induction, whereas only the $\bm B^\omega$ field of light can contribute to the spin induction in the absence of SOI. In both figures, we find that the induced spin polarizations are negligibly small when the SOI is absent. We again need to multiply by 30,000 or a larger number to make them visible in the present plot scales.

\begin{figure}
\includegraphics[
width=6cm]{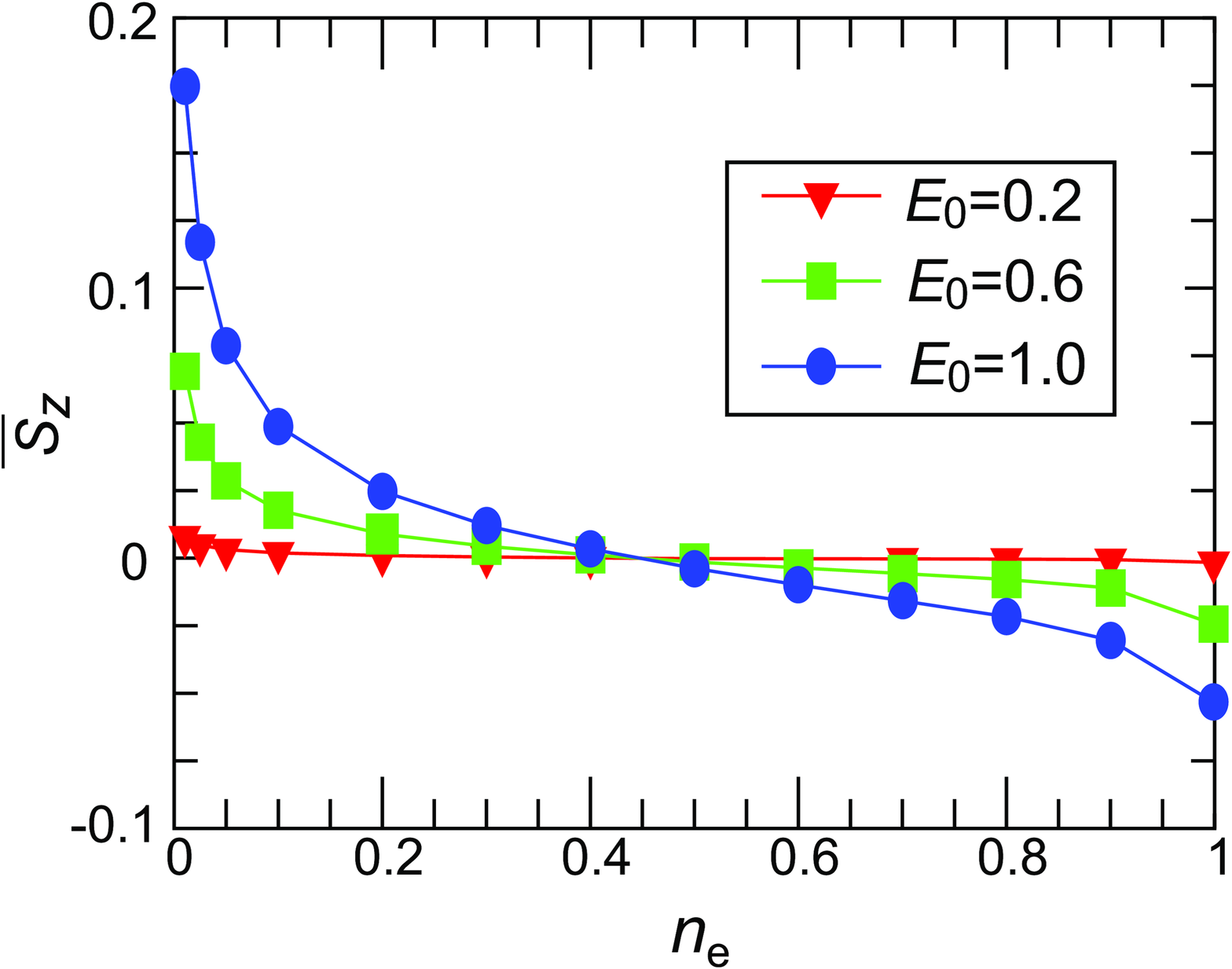}
\caption{Saturated spin polarizations $\overline{S_z}$ for several values of the amplitude of light electric field $E_0$ as functions of the electron filling $n_e$ when $\alpha_{\rm R}=0.1$ and $\omega=1$. See Table \ref{uc-table} for unit conversions.}
\label{Fig6}
\end{figure}
In Fig.~\ref{Fig6}, we show calculated filling dependence of $\overline{S_z}$ for several values of $E_0$ when $\alpha_{\rm R}=0.1$ and $\omega=1$. In the limit of small electron filling of $n_e\sim 0$, the spin polarization $|\overline{S_z}|$ takes a critically enhanced value. With increasing $n_e$ from $n_e\sim 0$, the value monotonically decreases toward $n_e=1$ traversing $\overline{S_z}=0$ near $n_e=0.45$. The characteristic behaviors of $\overline{S_z}$ with respect to $\omega$, $E_0$, and $n_e$ are accountable in terms of an effective magnetic field emerging from the SOI and the circularly polarized light as will be discussed in the next section.

\section{Floquet theory}
\begin{figure}
\includegraphics[
width=8cm]{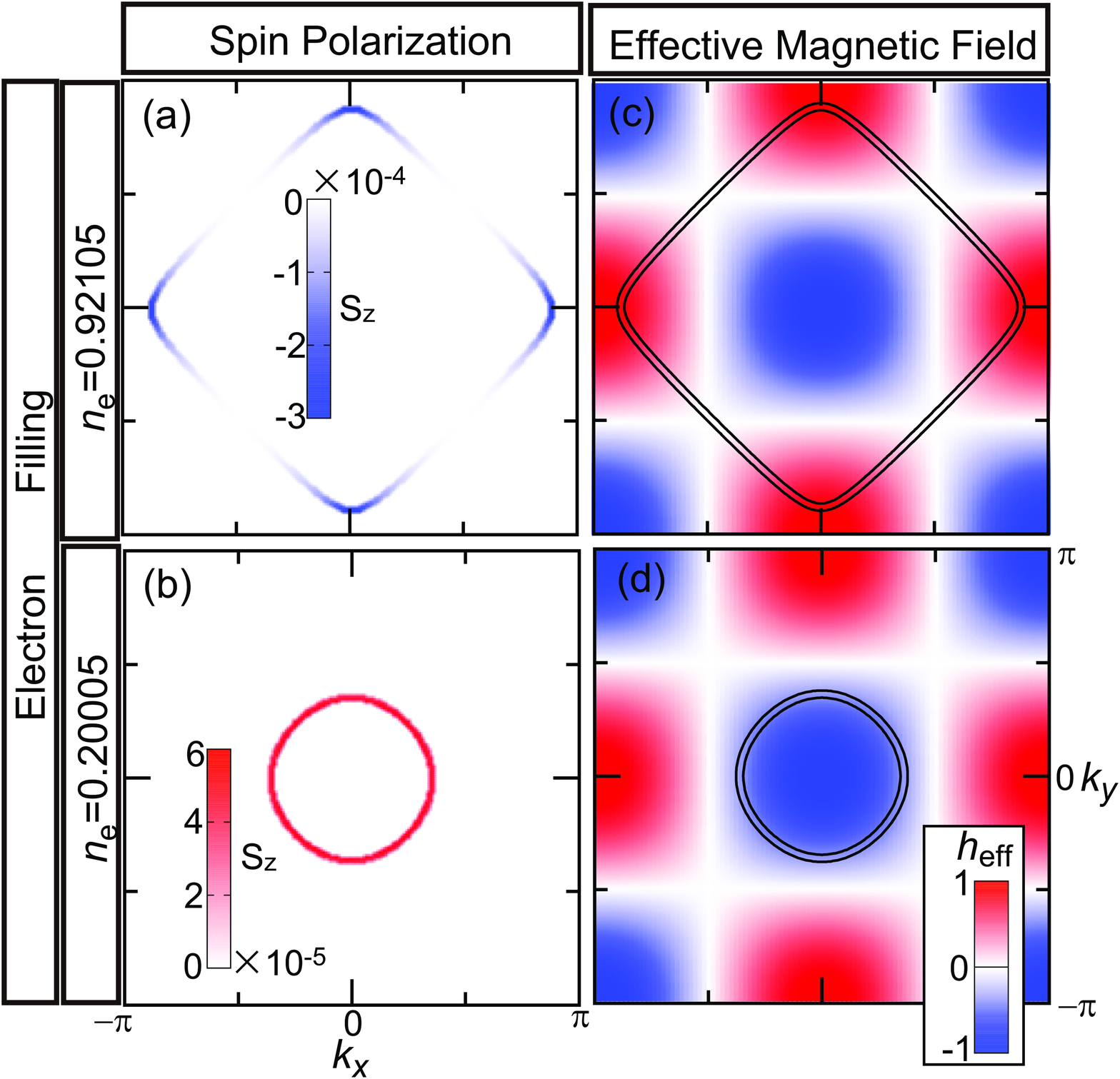}
\caption{(a), (b) Spin polarizations $s^z_{\bm k}$ in momentum space for electron fillings of (a) $n_{\rm e}$=0.92105 and (b) $n_{\rm e}$=0.20005 when $\alpha_{\rm R}$=0.1, $E_0$=0.5 and $\omega$=5. (c), (d) Momentum dependence of $h_{\rm eff}=h_{\rm eff}^0\cos k_x \cos k_y$ with $h_{\rm eff}^0=(\alpha_{\rm R} E_0)^2/\omega^3$ and Fermi surfaces for (c) $n_{\rm e}$=0.92105 and (d) $n_{\rm e}$=0.20005. Here $h_{\rm eff}^0$=$2\times10^{-5}$ for $\alpha_{\rm R}=0.1$, $E_0=0.5$ and $\omega=5$, which corresponds to 0.35 T when $t$=1 eV.}
\label{Fig7}
\end{figure}
We now construct an analytical theory based on the Floquet theorem~\cite{Kitagawa_PRB11,Lindner_NATP11}. The constructed theory turns out to describe not only the optical induction of spin polarization but also the peculiar laser-parameter dependence of the induced spin polarization discussed in the previous section. We derive an effective Hamiltonian, ${\mathcal H}_{\rm eff}$ for a system under application of continuous circularly polarized light field. Here we neglect the Zeeman coupling between the light magnetic field and the electron spins by setting $B_0$=0 or $b_\omega$=0 in the time-dependent Hamiltonian [Eq.~(\ref{eq-Hm})] because its contribution to the spin polarization has turned out to be several orders of magnitude smaller than that from the coupling between the light electric field and the electrons when the Rashba SOI is present. In the present Floquet theory, the time-dependent vector potential ${\bm A}(\tau)$ is written as
\begin{equation}
{\bm A}=(E_0/\omega)(-\sin\omega \tau, \cos\omega \tau).
\end{equation}
In the high-frequency limit with $\omega \gg t$ and  $\omega \gg \alpha_{\rm R}$, the effective Hamiltonian ${\mathcal H}_{\rm eff}$ can be derived from the formula~\cite{Kitagawa_PRB11,Lindner_NATP11,Eckardt_NJP15},
\begin{equation}
{\mathcal H}_{\rm eff}={\mathcal H}_{0}-\frac{1}{\omega}[
{\mathcal H}_{1},{\mathcal H}_{-1}], 
\end{equation}
where ${\mathcal H}_{m}$ ($m=0,\pm 1$) is defined by 
\begin{equation}
{\mathcal H}_{m}=\frac{1}{T}\int^{T}_{0}e^{im\omega \tau}{\mathcal H}(\tau)d\tau.
\end{equation}
Here $T(=2\pi/\omega)$ is the time period of the light. When $E_0/\omega \ll 1$, we can expand ${\mathcal H}_{\rm eff}$ with respect to $E_0/\omega$ as
\begin{equation}
{\mathcal H}_{\rm eff}=\sum_{\bm k}(c^\dagger_{\bm k\uparrow},c^\dagger_{\bm k\downarrow})
\begin{pmatrix}
\tilde{\varepsilon}_{\bm k,A}+h_{\rm eff}& \tilde{\gamma}_{\bm k,A} \\
\tilde{\gamma}^{\ast}_{\bm k,A} & \tilde{\varepsilon}_{\bm k,A}-h_{\rm eff}
\end{pmatrix}
\begin{pmatrix}
c_{\bm k\uparrow}  \\
c_{\bm k\downarrow}  
\end{pmatrix},
\label{eq-Hm-fl}
\end{equation}
where
\begin{eqnarray}
& &\tilde{\varepsilon}_{\bm k,A}=(1-\frac{E^2_0}{4\omega^2})\varepsilon_{\bm k}, \quad
\tilde{\gamma}_{\bm k,A}=(1-\frac{E^2_0}{4\omega^2})\gamma_{\bm k}, \\
& &h_{\rm eff}=-\frac{(\alpha_{\rm R} E_0)^2}{\omega^3}\cos (k_x)\cos (k_y).
\end{eqnarray}
In Eq.~(\ref{eq-Hm-fl}), it is apparent that the circularly polarized light effectively gives rise to a momentum-dependent {\it static} magnetic field, $h_{\rm eff}$, perpendicular to the plane of light polarization. In Figs.~\ref{Fig7}(a) and \ref{Fig7}(b), we show the spin polarization in the momentum space $s^z_{\bm k}$ for $n_{\rm e}$=0.92105 and $n_{\rm e}$=0.20005, respectively, when $\alpha_{\rm R}=0.1$, $E_0=0.5$ and $\omega=5$, which is given by,
\begin{eqnarray}
s^z_{\bm k}=\frac{1}{2}(\langle c^\dagger_{\bm k\uparrow}c_{\bm k\uparrow}\rangle-\langle c^\dagger_{\bm k\downarrow}c_{\bm k\downarrow}\rangle).
\end{eqnarray}
We find that finite values of $s^z_{\bm k}$ appear in the area sandwiched by the inner and outer Fermi surfaces. These color maps of $s^z_{\bm k}$ correspond to the distribution of $h_{\rm eff}$ in the momentum space shown in Figs.~\ref{Fig7}(c) and \ref{Fig7}(d). The appearance of finite $s^z_{\bm k}$ only in the area sandwiched by the split Fermi surfaces can be explained by an analytical formula of $s^z_{\bm k}$. By diagonalizing the effective Floquet Hamiltonian ${\mathcal H}_{\rm eff}$, we obtain eigenvalues $\widetilde{E}_{-}(\bm k)$ and $\widetilde{E}_{+}(\bm k)$ ($\widetilde{E}_{-}(\bm k)<\widetilde{E}_{+}(\bm k)$) and corresponding eigenvectors $(u_{-}(\bm k),v_{-}(\bm k))$ and $(u_{+}(\bm k),v_{+}(\bm k))$. The components of the eigenvectors are given by
\begin{eqnarray}
u_{\pm}(\bm k)&=&\frac{1}{\sqrt{2}}\frac{\tilde{\gamma}_{\bm k,A}}{|\tilde{\gamma}_{\bm k,A}|}
\Bigl(1\pm \frac{h_{\rm eff}}{\sqrt{h_{\rm eff}^2+|\tilde{\gamma}_{\bm k,A}|^2}}\Bigr)^{1/2}\\
v_{\pm}(\bm k)&=&\pm\frac{1}{\sqrt{2}}\Bigl(1\mp \frac{h_{\rm eff}}{\sqrt{h_{\rm eff}^2+|\tilde{\gamma}_{\bm k,A}|^2}}\Bigr)^{1/2}.
\end{eqnarray}
Subsequently, we obtain
\begin{equation}
s^z_{\bm k}=\frac{1}{2}(|u_{-}|^2-|v_{-}|^2)f(\widetilde{E}_{-})+\frac{1}{2}(|u_{+}|^2-|v_{+}|^2)f(\widetilde{E}_{+}), 
\end{equation}
where $f(\widetilde{E}_{\pm}) \equiv 1-\theta(\widetilde{E}_{\pm}-\widetilde{\epsilon}_F)$ is the Fermi distribution function at zero temperature. The Fermi energy for ${\mathcal H}_{\rm eff}$ is denoted by $\widetilde{\epsilon}_F$, and $\theta(x)$ is the Heaviside step function. We then have $s^z_{\bm k}=0$ when $f(\widetilde{E}_{-})=f(\widetilde{E}_{+})=1$, indicating that the spin polarizations in the upper and lower bands cancel each other. On the contrary, we have finite $s^z_{\bm k}$ given in the form
\begin{equation}
s^z_{\bm k}=-\frac{1}{2}\frac{h_{\rm eff}}{\sqrt{h_{\rm eff}^2+|\tilde{\gamma}_{\bm k,A}|^2}}
\end{equation}
when $f(\widetilde{E}_{-})=1$ and $f(\widetilde{E}_{+})=0$. For the higher electron filling of $n_{\rm e}$=0.92105, $|s^z_{\bm k}|$ is large near $(\pi,0)$ and $(0,\pi)$ in the $k$ space at which $|h_{\rm eff}|$ becomes large. Since $h_{\rm eff}$ is positive in this region, we have negative $s^z_{\bm k}$ and the resulting negative net spin polarization $S_z<0$. On the contrary, for the lower electron filling of $n_{\rm e}$=0.20005, the two Fermi surfaces are located in the region where $h_{\rm eff}$ is negative, which gives positive $s^z_{\bm k}$ and the positive net spin $S_z>0$. For $n_e\sim 0.45$, the Fermi surfaces are located at the intermediate position between the case of $n_e=0.20005$ and that of $n_e=0.92105$. In this case, positive and negative contributions to $\overline{S_z}$ due to $h_{\rm eff}$ cancel each other out and thus the spin polarization vanishes to be $\overline{S_z}=0$ as shown in Fig. \ref{Fig6}. In this way, the magnitude and sign of $S_z$ are determined by the momentum dependence of $h_{\rm eff}$ and the Fermi-surface geometry. 
Since the relation $h_{\rm eff}\propto E_0^2/\omega^3$ holds, the peculiar $E_0$-, $\omega$-, and $n_e$-dependencies of $\overline{S_z}$ observed in the numerical simulations are thoroughly explained by this Floquet analysis. 

Our results indicate that a smaller light frequency $\omega$ and a larger light amplitude $E_0$ are favorable to enhance the spin polarization $\overline{S_z}$. In addition, a smaller electron filling $n_e$ is preferable to generate a larger $\overline{S_z}$. It should be mentioned that the magnetic response to AC electric field in electron systems with SOI has been known as the electric dipole spin resonance (EDSR)~\cite{Rashba_SP60,Bell_PRL62,Melnikov_SP72} where the AC electric field acts on electron spins as an effective magnetic field via the SOI, which gives rise to the spin polarization. However, the momentum dependence of the effective magnetic field in this phenomenon has not been clarified, and the time evolution of the spin polarization is usually described by the Bloch equation which is formally equivalent to that in the electric paramagnetic resonance~\cite{Duckheim_NP06}.

\section{Discussion}
Now we discuss possible candidate materials relevant to our theoretical proposals. For small electron filling of $n_e$$\sim$ 0.01, relevant materials are n-type semiconductors with a large Rashba SOI due to their crystallographic structure with broken spatial inversion symmetry. A typical example is BiTeI in which the bottom of the conduction band is parabolic with a large Rashba splitting of $\sim 0.4$ eV~\cite{Ishizaka_NAM11}. A previous experimental study reported that the carrier density in BiTeI can be varied in a range from $0.2 \times 10^{19}$/cm$^3$ to $7.0 \times 10^{19}$/cm$^3$ by carrier doping~\cite{Lee_PRL11}. For the highest carrier density of 7.0 $\times$ $10^{19}$/cm$^3$, the Fermi level $E_{\rm F}$ is higher than $E_{\rm cross}$ by 0.1 eV where $E_{\rm cross}$ is the energy at the crossing point of two spin-split parabolic bands~\cite{Lee_PRL11}. The crossing point is located at the $\Gamma$ point for our model [see Fig. 1(b)]. In order to examine the applicability of our model to BiTeI, we evaluate the transfer integral $t$ by fitting the bottom of the conduction band. By using the relation $t\sim \frac{\hbar^2}{2m^{\ast}a^2}$, we obtain $t\sim 1.5$ eV where $m^{\ast}$ and $a$ are the effective mass of electron and the lattice constant, respectively. Here we use $m^{\ast}=0.1m$ ($m$ is the bare electron mass) and $a=5$ \AA, according to the experimental values for BiTeI~\cite{Ishizaka_NAM11}. For $n_e=0.01$ and $\alpha_{\rm R}=0.1$, our tight-binding model gives $E_{\rm F}-E_{\rm cross}=0.09$ eV, which coincides with the above-mentioned experimental result of $E_{\rm F}-E_{\rm cross}=0.1$ eV~\cite{Lee_PRL11}. These considerations support that our model is indeed relevant to BiTeI. We also note that for $\alpha_{\rm R}=0.1$ ($\alpha_{\rm R}=0.15$ eV for $t$=1.5 eV), the Rashba splitting is $\sim$ 0.14 eV. This value is smaller than the experimentally reported value of $0.4$ eV for BiTeI~\cite{Ishizaka_NAM11}, suggesting that the value of $\alpha_{\rm R}$ in BiTeI is, in reality, much larger than that used in our work, where we expect more prominent inverse Faraday effects. Therefore, BiTeI is one of the promising candidates to observe a large photoinduced spin polarization caused by the Rashba SOI.

On the other hand, when the electron filling is not small ($n_e$$\sim$0.1-1), relevant materials are metallic compounds with the Rashba SOI rather than semiconductors. One of the typical classes of materials is noncentrosymmetric superconductors Li$_2$Pd$_3$B and Li$_2$Pt$_3$B~\cite{Yuan_PRL06}, which are metallic above the superconducting critical temperature. In these compounds, the Fermi surface is split by the SOI~\cite{Lee_PRB05}. Noncentrosymmetric metallic compounds La$T$Ge$_3$ ($T$=Fe, Co, Rh, Ir) are another important class of materials, in which an observation of the Rashba-split Fermi surfaces was reported~\cite{Kawai_JPSJ08}. Although all these materials have a three dimensional crystal structure and their Fermi surfaces are more complex than those considered here, the proposed mechanism of light-induced spin polarization is essentially applicable to them.

We next discuss possible effects of spin relaxation. In real materials, relaxation of spins necessarily occurs~\cite{Elliott_PR54,Yafet_SSP63,Dyakonov_JETP71,Dyakonov_JETP72}, which may hinder the light-induced spin polarization. The relaxation time $\tau_s$ for semiconductors is typically in the range from 10 ps to 10 ns, although it depends on carrier density and temperature. In the case of simple metals, the typical spin-relaxation time is $\tau_s$=0.1 ns - 10 ns~\cite{Zutic_RMP04}. 
The spin relaxation time is governed by the scattering of electrons spins by phonons, magnons and magnetic impurities. Therefore, they are usually much longer than the time scale of electrons. Our results show that the spin polarization $S_z$ saturates within 100 fs (see Fig.~\ref{Fig2}). Thus, we expect that the spin polarization robustly occurs against the spin relaxation effects and thus can be detected experimentally. We consider that the pump-probe measurements of the magneto-optical Faraday effect~\cite{Kimel_NAT05} can be used to observe the proposed phenomenon. The peculiarity of our mechanism will manifest in the $n_e$ dependence of $\overline{S_z}$ shown in Fig.~\ref{Fig6}. For the n-type semiconductors such as BiTeI, this may be observed by electron-carrier doping since our results suggest that the spin polarization of photoinduced carriers steeply changes around $n_e=0$.

We also note that our numerical simulations were performed for a closed driven system without considering the effects of heating and dissipations, which inevitably occur in real experimental situations. However, the heating and dissipations have much longer time scale than the photoinduced spin plarization because they are caused by interactions between the present electron-spin system and the fluctuating environment of lattice and spin degrees of freedom such as phonons and magnons. The typical time scale of phonons is of the order of picoseconds, whereas that of magnons are nanoseconds or picoseconds, which are much longer than that of electrons. Our simulations for a typical parameter set of semiconductors demonstrated that the laser irradiation induces the ultrafast spin polarization, the process of which is finalized within a few hundred femtoseconds long before the heating and the dissipation set in. Thus our theoretical treatment is justified for the present study. On the other hand, it may be neccesary to consider the effects of heating and dissipations when we study the long-time dynamics of electron spins under persistent or continuous photo-irradiation~\cite{Haug_BOOK,Aoki_RMP14,Breuer_PRE00,Breuer_BOOK,Ikeda_arXiv}. 

\section{Summary}
To summarize, we investigated theoretically the inverse Faraday effect in the electron systems with the Rashba-type SOI by particularly foucsing on the effects of the formations of band dispersions and Fermi surfaces in the momentum space. Employing the tight-binding model with the Rashba SOI, we first performed the numerical simulations based on the time-dependent Schr\"odinger equation, which provide unbiased results for the spatiotemporal spin dynamics. The simulations demonstrated that the circularly polarized light induces the spin polarization perpendicular to its polarization plane, magnitude of which is proportional to $E_0^2/\omega^3$. We also found that the magnitude and even the sign of the induced spin polarization sensitively depend on the electron filling, which have been missed in the previous theoretical studies considering a continuum medium~\cite{Pitaevskii61} or an isolated ion~\cite{Pershan66}. The analytical theory that we constructed based on the Floquet theorem turned out to explain the observed $E_0$ and $\omega$ dependencies. The theory also attributed the observed sensitive filling dependence of the spin induction to the momentum-dependent effective magnetic field governed by the electron filling via the Fermi-surface geometry.

\section{Acknowledgment}
The authors would like to thank H. Munekata for fruitful discussions. This work was partly supported by JSPS KAKENHI (Grant Nos. 17H02924, 16H06345, 19H00864, 19K21858, 19K23427, 20K03841 and 20H00337) and Waseda University Grant for Special Research Projects (Project Nos. 2019C-253 and 2020C-269).

\end{document}